\documentclass[a4paper,fleqn,usenatbib]{mnras}
\usepackage{lineno}
\usepackage{blindtext}
\usepackage{newtxtext,newtxmath}
\usepackage[T1]{fontenc}
\usepackage{ae,aecompl}
\usepackage{epsfig}
\usepackage{amsmath}
\usepackage{graphicx}
\usepackage{url}
\usepackage{amssymb}
\usepackage{times}
\usepackage{CJK}

\usepackage{natbib}

\def\beq{\begin{equation}}
\def\eeq{\end{equation}}
\def\beqar{\begin{eqnarray}}
\def\eeqar{\end{eqnarray}}

\def\avg#1{\langle #1 \rangle}

\def\msol{M_\odot}

\def\isotope#1#2{\mbox{${}^{#2}{\rm #1}$}}
\def\fe5#1{\isotope{Fe}{5#1}}
\def\co5#1{\isotope{Co}{5#1}}
\def\ni5#1{\isotope{Ni}{5#1}}

\def\apj{ApJ}
\def\aap{A$\&$A}
\def\apjl{ApJL}
\def\apjs{ApJS}
\def\nat{Nature}

\def\mnras{MNRAS}
\def\prd{PhRvD}

\def\la{\mathrel{\mathpalette\fun <}}
\def\ga{\mathrel{\mathpalette\fun >}}
\def\fun#1#2{\lower3.6pt\vbox{\baselineskip0pt\lineskip.9pt
 \ialign{$\mathsurround=0pt#1\hfil##\hfil$\crcr#2\crcr\sim\crcr}}}
 
\title[Gamma Ray Alert for the Next Galactic SNIa]{Using Gamma Ray Monitoring to Avoid Missing
the Next Milky Way Type Ia Supernova}

\author[X. Wang et al.]{
Xilu Wang,$^{1,2,3}$\thanks{E-mail: xlwang811@gmail.com}
Brian D. Fields,$^{3,4}$
Amy Yarleen Lien$^{5,6}$
\\
$^{1}$Department of Physics, University of Notre Dame, Notre Dame, IN 46556, USA\\
$^{2}$Department of Physics, University of California, Berkeley, Berkeley CA, 94720, USA\\
$^{3}$Department of Astronomy, University of Illinois at Urbana-Champaign, Urbana, IL 61801, USA\\
$^{4}$Department of Physics, University of Illinois at Urbana-Champaign, Urbana, IL 61801, USA\\
$^{5}$Center for Research and Exploration in Space Science and Technology (CRESST) and NASA Goddard Space Flight Center, Greenbelt, MD 20771, USA\\
$^{6}$Department of Physics, University of Maryland, Baltimore County, Baltimore, MD 21250, USA\\
}

\begin{document}
\label{firstpage}
\pagerange{\pageref{firstpage}--\pageref{lastpage}}
\maketitle

\begin{abstract}
A Milky-Way Type Ia Supernova (SNIa) could be unidentified or even initially unnoticed, being dim in radio, X-rays, and neutrinos, and suffering large optical/IR extinction in the Galactic plane.   But SNIa emit nuclear gamma-ray lines from $^{56}{\rm Ni}\to ^{56}{\rm Co}\to ^{56}{\rm Fe}$ radioactive decays.  These lines fall within the {\it Fermi}/GBM energy range, and the \ni56 158 keV line is detectable by {\it Swift}/BAT. Both instruments frequently monitor the Galactic plane, which is transparent to gamma rays. Thus GBM and BAT are ideal Galactic SNIa early warning systems. 
We simulate SNIa MeV light curves and spectra to show that  GBM and BAT could confirm a Galactic SNIa explosion, followed by {\em Swift} localization and observation in X-rays and UVOIR band.  The time 
of detection depends sensitively on the \ni56 distribution, and can be as early as a few days 
if  $\ga$$10\%$ of the $^{56}{\rm Ni}$ is present in the surface as suggested by SN2014J gamma data.
\end{abstract}

\begin{keywords}
gamma-rays: stars -- (stars:) supernovae: general 
\end{keywords}


\section{Introduction} 
\label{sec:intro}

A Galactic SNIa explosion, i.e., one exploding in the Milky Way,  would uniquely advance our understanding in many ways. To name a few: a) The nature of Type Ia progenitors remains a mystery (single vs double degenerate scenarios). A Galactic SNIa would offer unique insight into the progenitor, e.g., via radio and X-ray probes of the circumstellar environment. b) SNIa are also the dominant sources of iron-group elements \citep{Nomoto}. Gamma-ray observations will precisely measure the $\ni56$ mass, and the lightcurves would encode a map of the structure and mixing of the ejecta. c) SNIa have been used as ''standardizable'' candles and thus cosmic distance indicators, famously leading to the discovery of cosmic acceleration. Multi-wavelength observations of a Galactic Type Ia event would offer a powerful new probe of the Phillips relation used to standardize the lightcurves \citep{Phillips}. Early detection strongly probes model differences and thus provides unique insight into all of these issues  \citep[e.g.,][]{Piro2013, Summa2013}

The Galactic SNIa rate per century has been estimated as $\sim0.51\pm0.12$ \citep{Li2011} or $1.4^{+1.4}_{-0.8}$ \citep{Adams2013}, 
making each explosion more rare than a ``once in a lifetime" event. Although SNIa are enormously luminous at peak, a Galactic event should lie in the plane of the Milky Way disk that contains obscuring dust.
Nonetheless, \citet{Adams2013} found that the deep optical/near-IR scanning of {\em Large Synoptic Survey Telescope (LSST) }can detect over $99\%$ of Galactic SNIa at peak in the brightest observable waveband, assuming {\em LSST} always monitors the Galactic plane.
 {\em LSST} will  thus play a crucial role in the event of a Galactic SNIa.
However, in practice one must not only {\em detect} but also {\em identify} the SNIa. Depending on its observed brightness, the outburst could initially be confused with less rare Galactic transients, like luminous red novae \citep[e.g.,][]{novae}, delaying the SNIa identification for $\sim$ 2 weeks or even longer.  This could push confirmation until after the optical peak, and the opportunity for early followup may be lost.\footnote{Several practical issues can delay the {\em LSST} identification until after the SNIa peak: 
(a) the {\em LSST} observing season (air mass <1.5), is only about 7 months per year (i.e., $60\%$) for the inner Galactic plane (\url {https://www.eso.org/sci/observing/tools/calendar/observability.html}); (b) even if the SNIa happens during the observing season, in the ''baseline" $minion\_1016$ \citep{LSST} observing strategy, {\em LSST will not visit the Galactic plane at all} beyond the first 7 months; (c) even the alternate proposed strategy astro$\_$lsst$\_01\_1004$ having Galactic plane scans throughout the {\em LSST} mission, the cadence is reduced to one visit every $\sim$ 10 days; (d) the explosion will occur randomly in the {\em LSST} filter cycle, and identification will involve comparing magnitudes and colors among filters that may not at first be optimal.} {\em LSST} very probably will discover a Galactic SNIa eventually, but having an independent means of identifying a SNIa increases the chances for early detection and maximal science returns.

Going beyond optical/IR, SNIa are faint in the radio and soft X-ray, as confirmed by the non-detection of the nearby SN2014J \citep{Perez-Torres2014, Margutti2014}. The SNIa neutrino signal is undetectable for events beyond $\sim$1 kpc \citep{Odrzywolek2011}. 
Gravitational wave observation of SNIa awaits space-based detectors \citep{Webbink}. Moreover, a future Galactic SNIa could happen anywhere anytime in the Galactic plane, most of which is not continually monitored at most wavelengths. 

Fortunately, SNIa are confirmed gamma-ray emitters, because the dominant product of these thermonuclear explosions is the radioisotope $^{56}{\rm Ni}$.  The decays $^{56}{\rm Ni}$$\xrightarrow{\rm 8.8 days}$$^{56}{\rm Co}$$\xrightarrow{\rm 111.3 days}$$^{56}{\rm Fe}$ emit gamma-ray lines spanning $158 \ {\rm keV}$ to $2.6 \ {\rm MeV}$. Observations of the Type Ia supernova SN2014J in M82 saw evidence for the dominant $^{56}{\rm Ni}$ lines at 158 keV and 812 keV within the first 20 days {\citep{Diehl2014, Diehl2015, Isern3}}. Later observations detected the $^{56}{\rm Co}$ lines at $847 {\rm keV}$ and $1238 {\rm keV}$ {\citep{Churazov2014, Churazov2015}}.   

The Milky Way is optically thin to gamma rays, so the \ni56 and \co56 lines provide a guaranteed signal for a SNIa anywhere in the Galaxy.
To exploit this signal as a 
SNIa alert requires $\sim$daily gamma-ray observations of the Galactic disk with a detector sensitive to the SNIa lines.
Fortunately, such data are taken as part of the ordinary operation of {\em Fermi}'s Gamma-ray Burst Monitor (GBM) and {\em the Neil Gehrels Swift Observatory (Swift)}'s Burst Alert Telescope (BAT). { We use these detectors to illustrate that the capability already exists, but future burst detectors and dedicated MeV missions (e.g., {\em AMEGO, e-ASTROGAM, LOX})
 can play the same role.}

{\it Fermi}/GBM consists of 12 NaI and 2 BGO scintillation detectors, with a large field of view (FOV) of $\sim$9.5 steradians {\citep{Meegan2009}} \footnote{\url{http://fermi.gsfc.nasa.gov}}. The NaI detectors are sensitive to 8 keV to 1 MeV photons, while BGO detectors work between 150 keV to 40 MeV, covering all the \ni56 and \co56 decay lines. {GBM/BGO has successfully identified solar flares in $e^\pm$ annihilation (511 keV) and neutron capture (2.22 MeV) lines \citep{solar}.}
{\it Swift}/BAT is a coded aperture telescope with a 1.4 steradian field of view {\citep[half coded;][]{Swift}}\footnote{\url{http://swift.gsfc.nasa.gov}}. The BAT energy range is $\sim$15-150 keV for imaging, which is just able to detect the \ni56 158 keV line. Typically, BAT views the Galactic center once per day and its field of view covers most of the probable regions where next Galactic SNIa will occur\footnote{If BAT is looking at the Galactic center, using BAT partial coding map (\url{https://swift.gsfc.nasa.gov/proposals/bat_cal/index.html}) and the probability distribution of Galactic SNIa \citep{Adams2013}, we estimate that $\sim$$77\%$ of the Galactic SNIa will fall within the FOV of BAT.}.
These properties make GBM and BAT ideal Galactic SNIa monitors and alarms. 

The next section shows the methods to confirm the viability of GBM and BAT
as SNIa detectors. Section~\ref{sec: results} presents the simulated SNIa gamma line signals seen by detectors and discussion of the results. In Section~\ref{sec:outlook}, outlook and recommendations are given.

\section{Methods}
\label{sec: methods}

To confirm the viability of GBM and BAT
as SNIa detectors, and assess the time needed to sound the alarm, we will 1) model the time history of the SNIa line emission, then locate our SNIa at a fiducial distance D$=$10 kpc\footnote{This distance is comparable to the distance to the Galactic Center, and consistent with the distance when the optical brightness is small for optical detection in \citet{Nakamura}. For the SNIa model in \cite{Adams2013}, the most probable distance is almost exactly 10 kpc, and $> 98$$\%$ of events are $<20$kpc. For other choices of distance, the flux and count rates will scale as $(10 {\rm kpc/D})^2$.},
and 2) simulate the line features (light curves and spectra) in GBM and BAT. 
With simulations of the SNIa spectra at different times (for light curves, see {Appendix~\ref{sec: lightcurves}}), we can estimate the timescale at which the rising SNIa signal emerges as distinct from the detector background. 
Strategies for localization and multimessenger followup appear in the {Appendix~\ref{sec: localization}}. 
Some early results of our analysis were summarized in \citep{wf}.
{In using burst monitors to search for Milky Way signals, we draw on pioneering blind searches for Galactic point sources \citep{Rodi} and diffuse emission \citep{ng}.}

 
 Our model of SNIa line emission follows the radioactive production of \ni56 and \co56 lines
 and their propagation in the expanding blast. The ejecta is initially dense and opaque,
 but becomes optically thin after $\sim$100 days 
 \citep{Sim2008, The2014}. {The gamma rays of interest lose significant energy in a single Compton scattering event,
so that lines in the emergent spectrum are only due to un-scattered photons.} 
Thus, the SNIa gamma line light curves are very sensitive to the structure of the \ni56 distribution in the ejecta.
In fact, the gamma line luminosity at early times ($\le$10 days) is dominated by surface emission set by the outermost velocity and \ni56 abundance. This early emission is characteristic of an expanding photosphere.

Later emission has been studied by a number of groups, where various SNIa models have been built with different \ni56 yields and distributions, ranging from interior-only \ni56 (e.g., deflagrations W7 model \citep{Nomoto}) to extra external \ni56 layers (e.g., \ni56 cap like delayed detonations W7DT model \citep{Yamaoka1992}, sub-Chandrasekhar helium detonations models \citep[][and references therein]{Hoeflich1996,Hoeflich2017}), or more detailed 3-D models \citep[e.g.,][]{Summa2013, Kromer}. The SNIa line fluxes from these models differ by $\sim$1-2 orders of magnitude at early time {\citep[the first 10s of days after the explosion;][and references therein]{Isern1, The2014}, showing that early gamma-ray observation will sharply probe SNIa models \citep[e.g., deflagration, detonation, delayed detonation and sub-Chandrasekhar models;][]{Isern1}.}

Our purpose is to evaluate the feasibility of gamma lines as tools for early SNIa detection, we therefore study two simple surface and interior \ni56 distributions that bracket an 
observationally-motivated optimistic case and the most pessimistic case: 
1) shell plus core model ({or helium cap model}, shortened as shell model thereafter), assuming total $^{56}{\rm Ni}$ mass $M_{^{56}\rm Ni,total}=0.5\msol$ with $10\%$ of the mass distributed at the outmost shell of the ejecta while the remain Ni is in the core; 2) core model, where all the Ni is distributed in the core. {For both models, the SNIa ejecta has uniform density, and expands homologously with total ejected kinetic energy $E_{\rm ej}=10^{51}$erg, and calculate lines only. The radius of the ejecta is $a=a_0+v_0t$, with $a_{0}\sim 7000$km is the initial ejecta radius, $v_{0}\sim0.04c$ is the velocity at the outermost radius. The total ejected mass is $M_{\rm ej}=1.4\msol$}. See { Appendix~\ref{sec:model}} for general equations and more details. 


{GBM and BAT are designed to identify transients like the gamma-ray bursts that exceed the background emission. A typical gamma-ray burst rise timescale is a few seconds, for which GBM and BAT triggering is optimized. But the gamma rays from a Galactic SNIa rise and decay on timescales of $\sim$ weeks ($t_{\rm mean, ^{56}{\rm Ni}}= 8.8 {\rm days}, t_{\rm mean, ^{56}{\rm Co}}=111.3 {\rm days}$), meaning the Galactic SNIa signal will appear as a long-duration increase in the background, instead of triggering the detector. So we need to compare our simulated spectra and light curves (light curves see {Appendix~\ref{sec: lightcurves}}) from a Galactic SNIa with the background of GBM and BAT detectors. In next section we find that a Galactic SNIa signal is indeed large enough to be noticed, and show how soon we can confirm the event after its explosion and sound the alarm. See section~\ref{subsec: discussion} for background variations}. 
 
In this paper, we adopt a rough criteria of {\em "detectable" to be when SNIa signal/background signal $\sim$1}, i.e., a "detected" Galactic SNIa point-source signal will be at least as large and likely larger than the entire all-sky background signal (see Section~\ref{subsec: discussion}). {\em In our analysis, "background" photons are all those seen by the detector in the absence of a SNIa, including both the instrument-induced noise and signals from diffusive emission and point sources.}

\begin{figure}
\centering
\includegraphics[width=0.95\columnwidth]{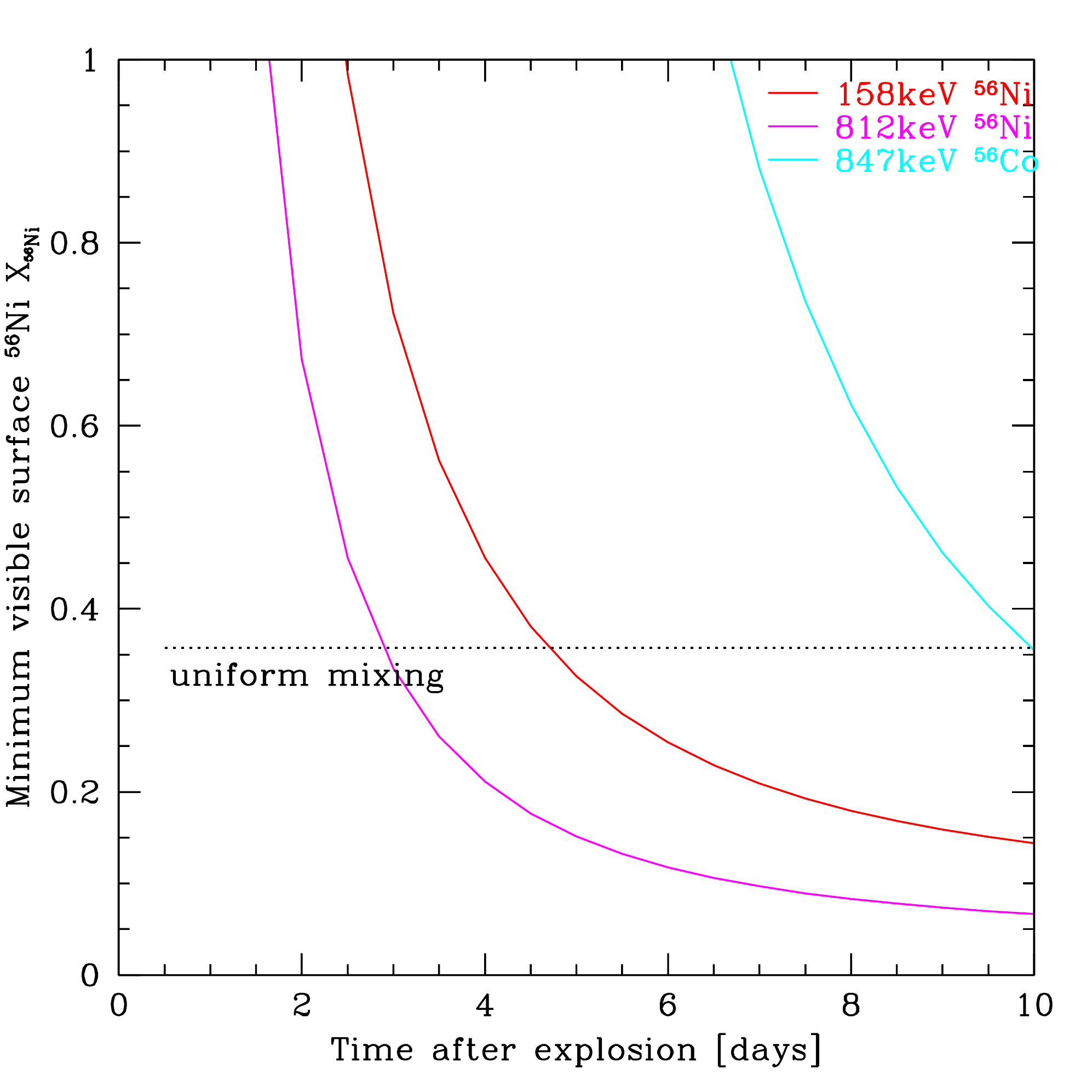}
\caption{
Minimum surface \ni56 abundance $X_{^{56}\rm Ni}$ versus the detected time for a 10 kpc Galactic SNIa.
\label{fig:x}
}
\end{figure}

Analytical formulas for SNIa flux can be obtained in the limits of optical depth $\tau_a$:

When the ejecta is {\em optically thick} ($\tau_a \gg 1$) at early time, we only see the photons coming from the surface ejecta which are facing us, thus the spectrum is a one-sided (blueshifted) linear rise with a sharp cutoff at $E_{\rm max}=E_i(1+v_0/c)$, where $E_i$ is the gamma ray line photon energy, $v_0$ is the surface velocity. This agrees with \citet{Bussard1989}'s calculation. The total line flux is thus:
\beq
\label{eq:f1}
F(E_i,t)=\pi S_i \frac{a^2}{D^2}=\frac{b_i X_{^{56}{\rm Ni}}}{4A_{^{56}{\rm Ni}}Y_e\sigma}\frac{(a_0+v_0t)^2}{D^2}f_{i}(t).
\eeq
which is {\em only determined} by the surface \ni56 abundance $X_{^{56}{\rm Ni}}$, electron fraction $Y_e$ and surface velocity $v_0$. Here $S_{i}$ is the source function, 
$b_i$ is the branching ratio of the gamma-ray line at $E_i$,  $\sigma$ is the {cross-section} of photons, $f_{i}(t)$ encodes the \ni56 and \co56 {decay} rates (details see {Appendix~\ref{sec:model}}).
This equation fits our simulation perfectly at early time, shown in the upper panel in Figure.\ref{fig:light_curve}.
Therefore for a certain SNIa, at early times when the ejecta is optically thick, there is an accurate analytical relation between $X_{^{56}{\rm Ni}}$ and the detected time $t_{\rm detected}$, defined as when the SNIa line signal is about the same as the detector background signal:
\beqar
\label{eq:X}
X_{^{56}{\rm Ni}}=\frac{A_{^{56}{\rm Ni}}Y_e\sigma}{b_i}\frac{4D^2}{(a_0+v_0t)^2}\frac{F_{\rm background}(E_i,t_{\rm detected})}{f_{i}(t_{\rm detected})}
\eeqar
which is shown in Fig.~\ref{fig:x} for GBM/BGO detection. For a ``uniform mixing" scenario, i.e., \ni56 is uniformly distributed in the spherical ejecta with $X_{^{56}{\rm Ni}}\sim$$35.7\%$, the \ni56 158keV signal can be seen as early as $\sim$day 3. So a SNIa signal is able to be seen at early time, as long as the surface \ni56 abundance is at least a few percent, and may be distributed in any shape, e.g., a ``plume".

When the ejecta is {\em optically thin} ($\tau_a\ll 1$) at late time, {we see the entire remnant (nebular phase), the spectrum is an inverted parabola centered on $E_i$ with endpoints at the extrema $E_\pm=E_i(1\pm v_0/c)$, the shape is close to a gaussian near the peak (expansion to the 2nd order).} Then the total line flux is
\beqar
F(E_i,t)=\frac{b_i}{4\pi D^2}\frac{dN_{\rm Ni, or, Co}}{dt}=\frac{b_i}{4\pi D^2}\frac{M_{Ni}}{A_{^{56}\rm Ni}m_p}f_{i}(t),
\eeqar
which only depends on the total \ni56 mass $M_{^{56}{\rm Ni}}$ of the SNIa. 
Thus for a certain SNIa, with whatever ejecta distribution, the line flux is always the same when the ejecta is optically thin.

\section{Results}
\label{sec: results}

In our analysis, both the GBM and BAT detectors' background and response files are from HEASARC site, processed with {\tt Xspec}
\footnote{\url {http://heasarc.gsfc.nasa.gov/xanadu/xspec/,
http://heasarc.gsfc.nasa.gov/FTP/fermi/data/gbm/, ftp://legacy.gsfc.nasa.gov/swift/calib_data/bat/}}. 
Background files are selected from typical GBM (cspec files) and BAT daily files for a particular detector when the satellite is looking at the Galactic center.
{For BAT, we simulate the response of a point source signal taking into account of the coded mask effect. For GBM analysis, we simulate the signals for individual detector (BGO: b0 and b1; NaI: n0 to n9, na, nb). Here we show the simulation results for detector b1 and n3 as examples.}

\subsection{Spectrum}
\label{sec: spectrum}

\begin{figure}
\centering
\includegraphics[width=0.95\columnwidth]{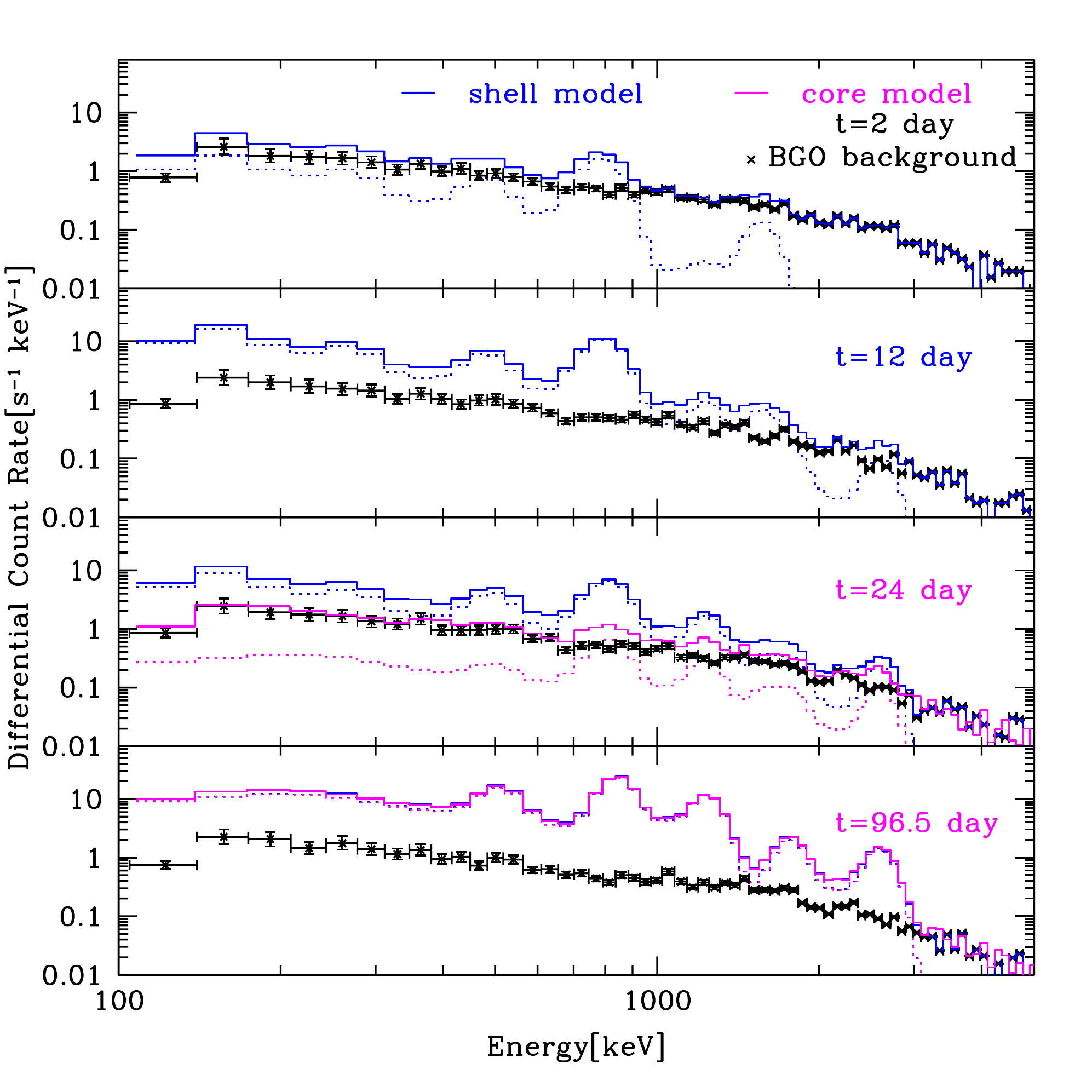}
\caption{
Simulation of $^{56}{\rm Ni}$ and $^{56}{\rm Co}$ gamma-ray lines 
signal seen in a GBM/BGO detector, from a 10 kpc Galactic SNIa at day2 (above), day12 (middle above), day24 (middle below) and day96.5 (below) after explosion. The black points represent a typical background spectrum from detector b1. Dotted lines are the simulated SNIa \ni56 and \co 56 decay lines signal, and solid lines are the total signal expected to be seen (sum of the background and SNIa signal). Blue lines are for shell model, magenta are for core model.   
\label{fig:BGO_spectra}
}
\end{figure}

\begin{figure}
\centering
\includegraphics[width=0.95\columnwidth]{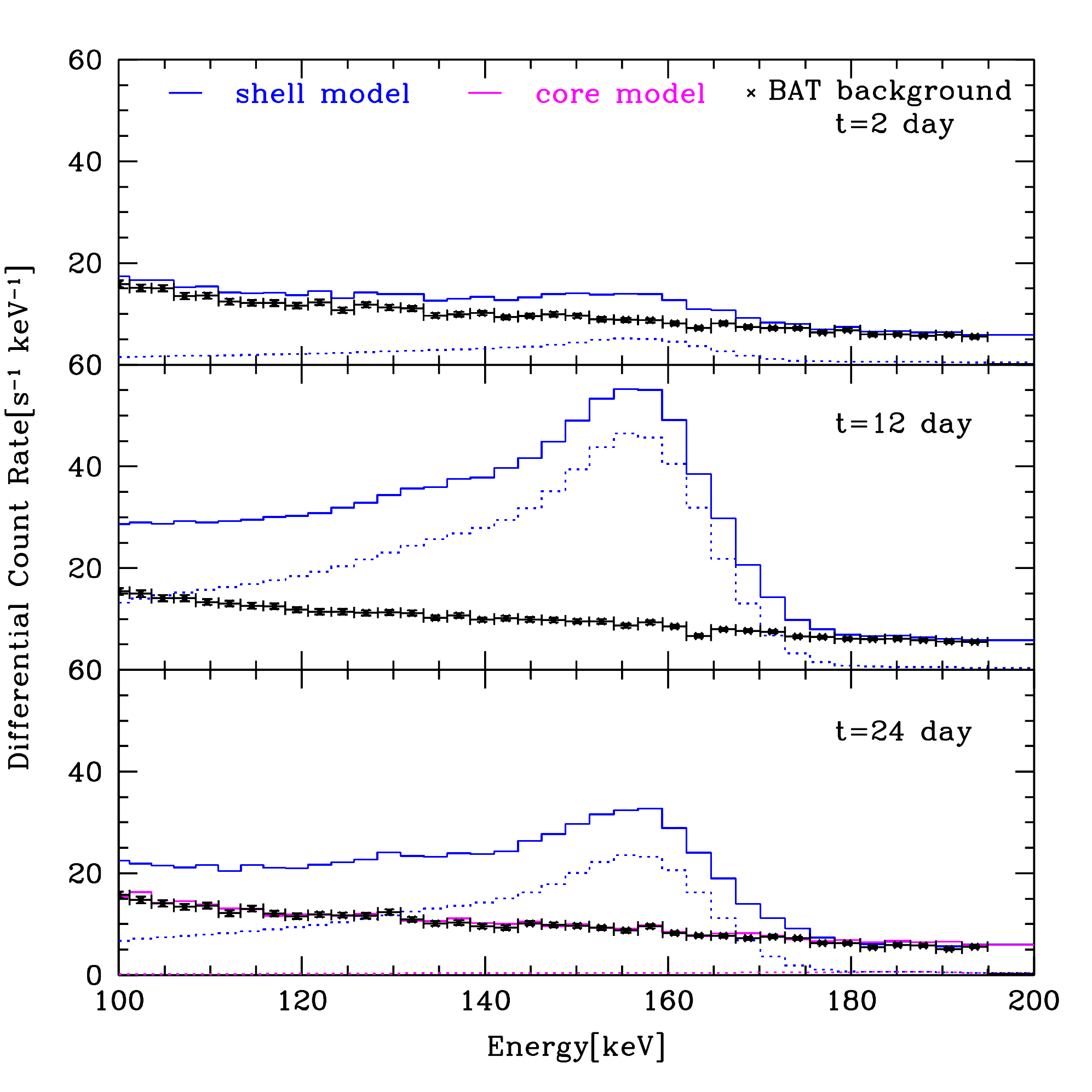}
\caption{Similar to Fig.~\ref{fig:BGO_spectra}, simulated gamma signal seen on-axis in BAT detectors, black points here represent a typical background spectrum from all the BAT detectors.
\label{fig:BAT_spectra}
}
\end{figure}

{\em The SNIa signal will be distinct from other high-energy transient events in the Galactic plane due to
the presence of the decay lines, which serve as the SNIa ``smoking gun.''}  Thus, for the two models described in Section~\ref{sec: methods}, we simulate the spectra of $^{56}{\rm Ni}$ and $^{56}{\rm Co}$ decay lines from a Galactic SNIa in a single BGO detector b1 ({other detectors will have the similar results with different phases}), for times day 2, day 12 (when \ni56 lines peak), day 24 (when \co56 lines emerge for core model) and day 96.5 (when \co56 lines peak) after the explosion, shown in Fig.~\ref{fig:BGO_spectra}. {We assume the lines are Doppler broadened corresponding to a line-of-sight velocity of 
0.04c, which resembles a gaussian shape at late time}. {For alert purpose at early days after explosion, we only use the gamma lines from the un-scattered photons to compare with the background to determine whether and when a SNIa can be detected. The Compton-scattered continuum changes little of the line profile, thus is neglected here, but see Section~\ref{subsec: discussion} for more discussion.} 

Fig.~\ref{fig:BGO_spectra} shows a time series of spectra for the optimistic shell model and pessimistic core model of \ni56 ejecta. SNIa $^{56}{\rm Ni}$ lines start to be detectable by GBM ($\gtrsim 100\%$ of the BGO background) at just 2 days after the explosion. 
By 12 days, all of the $^{56}{\rm Ni}$ lines are distinct, and are $\sim$3-4 times bigger than the typical background; $^{56}{\rm Co}$ lines (especially the dominant 847 keV line) begin to arise as well. Thus, in the shell model we expect GBM could detect a Galactic supernova within a few days, with spectral line features from both $^{56}{\rm Ni}$ and $^{56}{\rm Co}$ serving to confirm the Type Ia origin of the explosion. 
 
By contrast, the core model lacks surface emission, instead placing the nickel maximally deep.  In this pessimistic scenario,
the early ejecta are opaque and $^{56}{\rm Ni}$ lines are absent. Fig.~\ref{fig:BGO_spectra} shows that $^{56}{\rm Co}$ lines start to emerge above background around day 24, and rise to peak at $\sim$100 days. Therefore, in the core model, we expect GBM could only observe $^{56}{\rm Co}$ lines to detect a Galactic SNIa, starting around 24 days.
Similar spectra and conclusions are also obtained for the NaI detectors {of GBM}, the spectra plots for detector n3 are shown in Figure.~\ref{fig:NaI_spectra}.

We also simulate the spectra in BAT detectors at day 2, day 12 and day 24 in Fig.~\ref{fig:BAT_spectra}. BAT will not be able to detect $^{56}{\rm Ni}$ 158 keV line for the core model as the ejecta is optically thick at early phase after the explosion. For the shell model, the on-axis signal amplitude of $^{56}{\rm Ni}$ 158 keV line is $\sim$4-5 times larger than the average BAT background flux at day 12 after the explosion. 
{At day 2, the 158 keV signal is weaker than the background, but it may be detectable then -- or even earlier -- via the more sensitive image trigger technique \citep{Krimm}\footnote{The formal significance of this day 2 signal in the 140-180 keV band is $\sim30$ sigma in 100s' observation, adopting standard analysis method \citep{Li&Ma}, and the formal signal significance increases to $\sim90$ sigma with the 5$\sigma$ sensitivity limit of BAT imaging (see section.~\ref{subsec: discussion}). We thank the referee for pointing this out.}.

To conclude, if a SNIa were to explode in our Galaxy, {\em Fermi}/GBM will be able to detect the signal as early as $\sim$day 2 after the explosion if there is significant surface nickel, or no later than $\sim$day 24 for a pessimistic core model case. In Section~\ref{sec: methods} we show that even a surface mass fraction $X(\ni56)\sim$$10\%$ allows detection within $\sim$10 days. {\em Swift}/BAT can only see the $^{56}{\rm Ni}$ 158 keV line, and thus requires surface nickel for detection.  
Moreover, if GBM see SNIa lines other than 158 keV, but BAT sees no line signal, this means that there is no surface \ni56 in the SN ejecta.  

A SNIa detection would triggers a host of multi-wavelength observations.  Speed and reliability of the alarm will be crucial.
There are two possibilities of getting alert of a Galactic SNIa: 1) BAT discovers the 
SNIa first, or 2) {\em Fermi} finds the SNIa first and use the Earth occultation technique \citep{Wilson-Hodge2012, Rodi} to localize it within degrees. After the alert by BAT or {\em Fermi}, {\em Swift} will slew to the SNIa, localize it within arcminutes, and take XRT and UVOT spectra. For details of GBM and BAT localization, see {Appendix~\ref{sec: localization}}.

\subsection{Discussion}
\label{subsec: discussion}

As the two BGO detectors and 12 NaI detectors mounted on GBM are pointing at different direction, if a SNIa signal is seen in one detector, other detectors are expected to observe the similar signal with phase delay. 
BAT background file is the total background signal measured by all the active detectors. We have included both the mask effect and the detector degrading effect. For an on-axis signal, half of the detectors are masked by the coded aperture mask. And the active detectors of BAT are decreasing with time \citep{Lien}, therefore we use recent background signal (2017 year) and $50\%$ as the active detector percentage (current active detector ratio is $\sim$$56\%$) for a fair comparison. 
When BAT is looking at the Galactic center, its FOV is $\pm$ 20 degree along the long axis. For our simulations, the 26.56 deg off-axis signal is $\sim$$90\%$ of the on-axis signal from the same Galactic SNIa, thus we estimate the off-axis effect will decrease $\sim$$10\%$ of the supernova signal.

Any supernova signal in the GBM and BAT must compete with the (time varying) gamma-rays background that always presents. 
Above $\sim$150 keV, the background sources are dominantly secondary gamma rays created by cosmic-ray interactions in the Earth's atmosphere, and in the spacecraft itself \citep{Meegan2009}. 
The average variations of both GBM and BAT background over one orbit are about $\sim$$50\%$ \citep{GBM, BAT}, due to the variation of cosmic-ray flux densities in the atmosphere. 
Although the background is dominated by cosmic rays, which are affected by solar activity, we find that the variations over years or solar cycle is quite small ($\lesssim$$20\%$), based on our crude analysis of the detectors' daily data from different years. 
The background variation analysis here exclude orbits passing the South Atlantic Anomaly (SAA). 
These background variations can be seen as a noise source that is much larger than, and in addition to, the detectors' background errors in each exposure due to counting statistics.
Therefore we require that SNIa signal/background $\sim$1 for discovery or detection, which 
we believe is a good first approximation and a safe, conservative approach when one does not know the background level at the time of the possible discovery. 
 
Accounting for all these effects, the SNIa signal will still be detectable by both GBM and BAT within several days after explosion if there is significant surface \ni56 in the SNIa. If there is not, we will have to wait weeks. The uncertainty of surface \ni56 abundance dwarfs other concerns.

It is likely that more sensitive SNIa line search methods are possible and could lead to even earlier detection.
For GBM, every 15 or 30 orbits the satellite flies over the same location on Earth, so background subtraction is more sensitive at these intervals.
Even then there are worries about variations in photomultiplier gain \citep{GBM}.  
BAT will perhaps do a better job for early signal detection, especially using the image trigger technique \citep{Krimm}. For example, {with an exposure time of $\sim10^3$s, the 5$\sigma$ noise level, achieved in the BAT 70 month survey from the all-sky mosaic maps, is $\sim$ $1.6\times10^{-3}{\rm counts}\ {\rm s}^{-1}{\rm detector}^{-1}$ in the 14-195 keV range, or $\sim$ $10^{-2}{\rm s}^{-1}{\rm cm}^{-2}$ in the 146.2-170.1 keV bins which contain the broadened 158keV line} \citep{BAT2},
which is far smaller than the SNIa $158 {\rm keV}$ signal at first day. {
Even for SNIa without surface \ni56, the gamma flux in the 50-150 keV band is $\sim0.1{\rm s}^{-1}{\rm cm}^{-2}$ at day 8 \citep{Isern2}, possibly detectable by BAT with this technique within the first week after the explosion. 
However, the potential detection difficulties are the imaging and localization limit of BAT around 150-200 keV.}

We only consider the nuclear lines emitted from SNIa (zero-scattered photons from the $^{56}{\rm Ni}$ and $^{56}{\rm Co}$ decay) here for a conservative analysis of the SNIa observation. The Doppler broadening of the decay lines is below the resolution of {\em Fermi} and {\em Swift}.
In addition, the instrumental response introduces "fake" continuum in the simulated spectra, dominant over the true continuum due to Compton scattering (the simulation of detected lines are much wider than the theoretical lines). The signal will be enlarged if the Compton continuum is included (adding scattered photons), making the signal easier to detect. 
Partial energy deposition in the detector also leads to an instrumental low-energy shelf below the lines, e.g., Fig. ~\ref{fig:BGO_spectra}, 100-400 MeV.

Based on our simulations of SNIa spectra and light curves, even for an extreme distance of 20 kpc, the signal will eventually be detectable (e.g., at $\sim$ day 12 for the shell model).  Thus
essentially any Galactic SNIa will be detectable by {\em Fermi} and {\em Swift}, or future gamma ray telescopes with no worse sensitivity.  Obviously, the
closer the event, the sooner its gamma signature will emerge.

\section{Outlook and Recommendations}
\label{sec:outlook}

Although a Galactic SNIa is rare on human timescales, the potential scientific impact merits preparations--a similar philosophy was recently vindicated by the spectacular GW170817 \citep{GW}. Luckily, our analysis shows {that {\em Fermi}/GBM and {\em Swift}/BAT already are capable of sounding the alarm and welcoming the next Galactic SNIa, {\em without need for modifications in observing strategies}. 

We recommend the following to prepare for this inevitable and exciting event.
(a) We strongly urge that {\em LSST} scans the Galactic Plane over the entire mission duration, ideally with the same cadence as the main wide-fast-deep survey \citep{Strader}.
(b) We recommend that future gamma-ray burst missions succeed existing missions without gaps in time, do not avoid the Galactic Plane in the scans, and ideally are sensitive to {the energy range $\sim50\ {\rm keV} - 2\ {\rm MeV}$}; the SNIa search can piggyback on the GRB-focused mission without additional cost.
(c) MeV telescopes with large fields of view like {\em AMEGO} or {\em e-ASTROGAM} or {\em LOX} will be ideal \citep{MeV}, seeing all of the brightest \ni56 and \co56 lines, and may even resolve line widths and track them over time, testing predictions by several groups \citep[e.g.,][]{The2014} and probing the ejecta dynamics. The 511keV line, which maps the positron annihilation and thus separately probes density, will also be detectable; if there is sensitivity to the gamma line polarization, this will give even more information of SNIa \citep{polarization}.  Such missions will likely detect extragalactic SNIa as well. 
(d) A {\it Swift} like X-ray and UVOT combination would be very useful for rapid localization and initial multiwavelength observations.}

{There still remains room to improve our work. We are now working on the Monte Carlo calculations for various models with different ejecta structures, and including the continuum emission, to simulate a range of the SNIa signals. 
Other future work includes building pipeline codes for {\em Fermi} and {\em Swift} SNIa detections with better detection criteria, and simulating SNIa signals for future MeV telescopes like {\em AMEGO}.}

\section*{Acknowledgements}

BDF thanks Raph Hix for conversations that inspired this work. We are also pleased to acknowledge fruitful conversations with John Beacom, Valerie Connaughton, Ryan Foley, Athol Kemball, Sylvain Guiriec, Shunsaku Horiuchi, Kenny C.Y. Ng,  and Paul Ricker. This work was supported in part by the NASA Swift GI program, grant NNX16AN81G. This work also benefited from discussions at the 2018 Frontiers in Nuclear Astrophysics Conference supported by the National Science Foundation under Grant No. PHY-1430152 (JINA Center for the Evolution of the Elements).

\bibliographystyle{mnras}
\bibliography{ref_SNIa}


\appendix

\section{Model} 
\label{sec:model}

The procedure to conduct the research is as following. First, build models for a SNIa's ejecta to calculate the curves and spectra of both $^{56}{\rm Ni}$ and $^{56}{\rm Co}$ decay from a Galactic SNIa. Then simulate what the light-curves and spectra would look like in GBM and BAT detectors, respectively. Thirdly, compare the simulated Galactic SNIa signal with the typical background of the detector to see whether the signal is large enough to be noticed at early days and how soon we can confirm the signal is from a Galactic SNIa. Finally, after the confirmation of the signal, localize the SNIa with good accuracy in a short time using {\em Fermi} and {\em Swift}, and the followup multi-wavelength and multi-messenger observations.

Here we build models that explore both the optimistic and pessimistic case of ejecta scenarios (shell plus core model and core only model). These will serve as inputs to the radiation transfer calculations, yielding the SNIa $\gamma$-ray light curves and spectra. {The fiducial distance is set to be D =10 kpc}.

To estimate the light curve and spectrum requires a model for the gamma-ray emission and transfer in the ejecta.  
Previous work made the assumption (plausible at the time) that SNIa are spherically symmetric and stratified, with \ni56 buried deep in the core \citep[e.g.,][]{Bussard1989}. These calculations found that SNIa are opaque to gamma-rays for $\sim$100 days due largely to Compton scattering in the initially dense ejecta.  If this were the case, then a Galactic SNIa would likely not be discovered until months after the explosion, delaying followup observations until well after the optical peak emission.
Fortunately, {\em INTEGRAL} observations of SN2014J reported $^{56}{\rm Ni}$ lines within $\sim$20 days after the explosion, far earlier than expected.  This initial line flux corresponds to about $10\%$ of the total expected $^{56}{\rm Ni}$ mass \citep{Diehl2014,Diehl2015} at the surface (in fact, the proposed geometry has the nickel concentrated in a belt).
Later observations detected the \co56 lines which imply an initial \ni56 mass very close to $0.5 M_\odot$ \citep{Churazov2014,Churazov2015}.

We adopt a zeroth-order uniform ejecta model following \citet{Bussard1989}'s work, where the electron number density profile $n_{\rm e}$ is assumed to be flat in radius and drops to zero at the outer radius of ejecta $a=a_0+v_0t$. Here $a_{0}$ is the ejecta radius at the explosion time, $v_{0}$ is the velocity at the outermost radius and is given by $v_{0}=(10E_{\rm ej}/3M_{\rm ej})^{1/2}\sim1.3\times10^{7}m/s\sim0.04c$, $M_{\rm ej}$ is the total mass of the debris, $E_{\rm ej}$ is the total kinetic energy ejected. Therefore the volume of the ejecta is $V=4\pi a^3/3$, $n_{\rm e}={Y_{\rm e}}n_{\rm b}=M_{\rm ej}/({\mu_{\rm e}}m_{\rm p}V)$, where $n_{\rm b}$ is the baryon density, $Y_{\rm e}$ is the mean electron number per baryon, $m_{\rm p}$ is the proton mass.{ We adopt $M_{\rm ej}\sim M_{\rm Chandrasehkhar}=1.4\msol$,  $E_{\rm ej}=10^{51}$erg, $a_0=r_{\rm white-dwarf}\sim 7000$km, and $Y_{\rm e} \approx1/2$ here for a Galactic SNIa.}

\begin{figure}
\centering
\includegraphics[width=0.95\columnwidth]{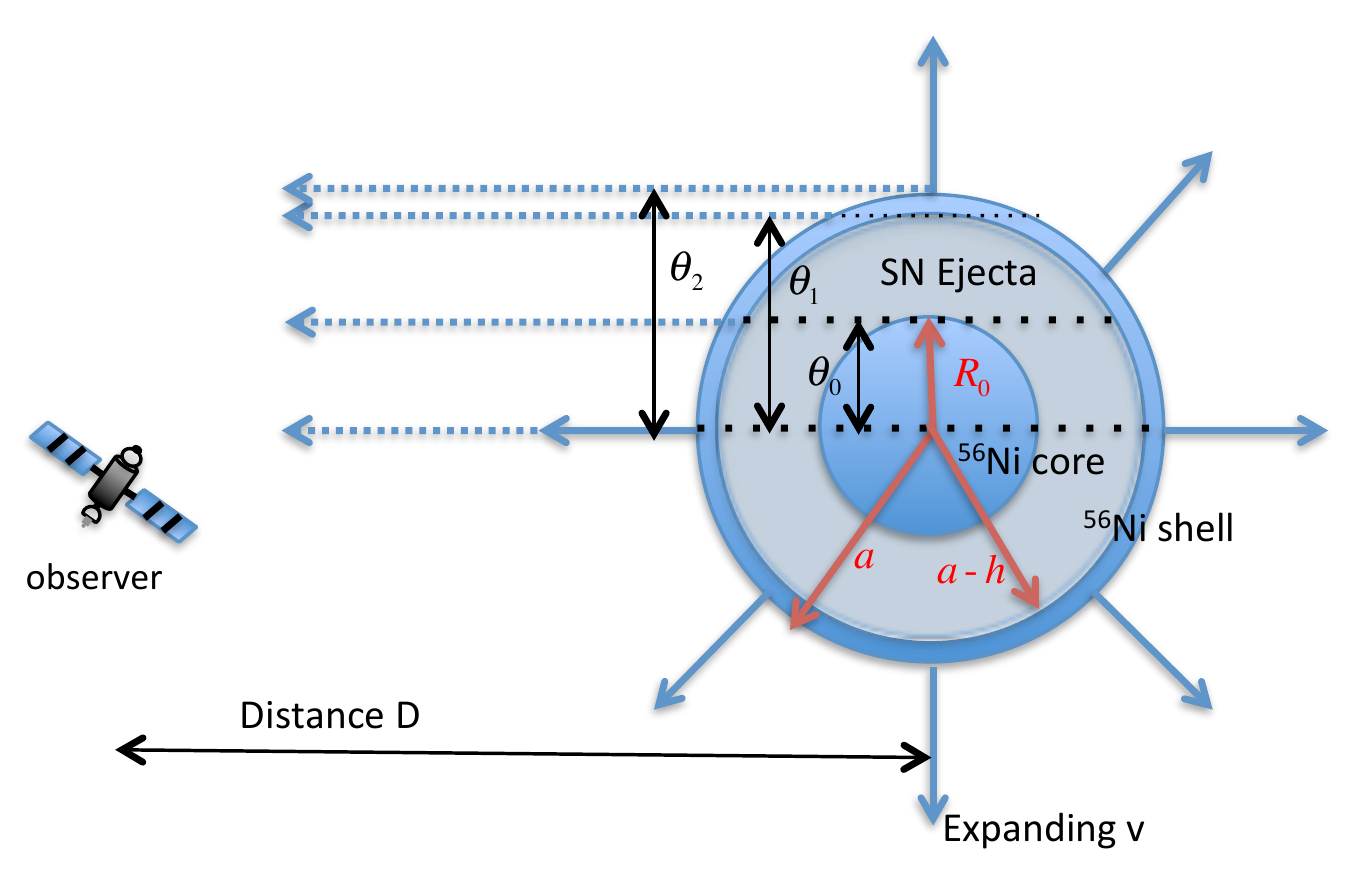}
\caption{
Sketch of the shell plus core model (shell model). For this model, $^{56}{\rm Ni}$ is distributed at both the ejecta's outmost shell ($10\%$ of the mass) with depth h, and the inside core with radius R$_0$. The ejecta radius is a, the distance between the detector and the SNIa is D.
\label{fig:shell}
}
\end{figure}

For uniform density ejecta, the mass density is just $\rho=\rho_{\rm Ni}=M_{\rm ej}/V_{\rm ej}$, therefore $n_{\rm Ni}/n_{\rm e}=(\rho_{\rm Ni}/m_{\rm Ni})/(Y_{\rm e}\rho/m_{\rm p})=1/Y_{\rm e}A_{^{56}{\rm Ni}}$.
The ejecta is expanding homologously, i.e., $v(r)\propto r$, r is the radius from the center $r\sim vt$.

The number emission coefficient for line $E_i$ at energy E is 
\beqar
\label{eq:coeff}
j_{E} & = & \frac{dN_i}{dtdVd\Omega} \phi(E)=b_i\frac{dN_{\rm Ni, or, Co}}{dtdVd\Omega}\phi(E)
\nonumber\\
&=&b_i\frac{dn_{\rm Ni, or, Co}}{dtd\Omega}\phi(E)=j_0\phi(E)
\eeqar
where $j_0$ is the emissivity when the line is not broadened, $b_i$ is the branching ratio of the gamma-ray line at photon energy $E_i$. {$\phi(E)$ is the Doppler effect term, $\phi(E)=\delta(E-\epsilon)$, where $\epsilon=E_i[1+cos\alpha v(r)/c]$ ,with $\alpha$ the angle between the radius vector and the line of sight.} This doppler effect term will broaden the decay lines but at a level below the resolution of {\em Fermi} and {\em Swift}.

The source function is 
\beqar
\label{eq:s}
S_{E} =\frac{j_{E} }{n_{\rm e}\sigma}=\frac{b_i}{n_{\rm e}\sigma(E_i)}\frac{dn}{dtd\Omega}  \phi(E)=\frac{b_i}{4\pi n_{\rm e}\sigma(E_i)}\frac{dn}{dt}  \phi(E)
\eeqar
where, 
\beqar
\label{dn/dt}
\frac{dn_{\rm Ni}}{dt}|_{\rm decay}&=&\frac{n_{\rm Ni}|_{\rm decay}}{t_{\rm mean, ^{56}{\rm Ni}}}=\frac{n_{\rm Ni}|_{\rm decay}(t=0)}{t_{\rm mean, ^{56}{\rm Ni}}}e^{-t/t_{\rm mean, ^{56}{\rm Ni}}}
\nonumber\\
&=&n_{\rm Ni}|_{\rm decay}(t=0) f_{\rm Ni}(t)
\eeqar
\beqar
\frac{dn_{\rm Co}}{dt}|_{\rm decay}&=&\frac{n_{\rm Ni}|_{\rm decay}(t=0)}{t_{\rm mean, ^{56}{\rm Co}}-t_{\rm mean, ^{56}{\rm Ni}}}(e^{-t/t_{\rm mean,^{56}{\rm Co}}}
\nonumber\\
& &-e^{-t/t_{\rm mean, ^{56}{\rm Ni}}})
\nonumber\\
&=&n_{\rm Ni}|_{\rm decay}(t=0) f_{\rm Co}(t).
\eeqar
To avoid confusion with optical depth $\tau$, we use $t_{\rm mean, ^{56}{\rm Ni}}= 8.8 {\rm days}, t_{\rm mean, ^{56}{\rm Co}}=111.3 {\rm days}$ for the mean lifetime of the radioactive $\ni56$ and $^{56}{\rm Co}$, respectively. $\sigma(E_i)$ is the cross-section when photons propagate through the dense material of the ejecta (use \co56 as an estimation), including {Rayleigh scattering}, {Compton scattering} ({Klein Nishina cross section}), photoelectric absorption, and pair production, values adopted from XCOM website\footnote{\scriptsize{\url{https://www.nist.gov/pml/xcom-photon-cross-sections-database}}}.

Therefore the source function can be written as 
\beqar
\label{eq:s1}
S_{E}=b_i\frac{X_{^{56}{\rm Ni}}}{4\pi \sigma A_{^{56}{\rm Ni}}Y_{e}} f_{\rm Ni\ or\ Co}(t)  \phi(E)=S_0\phi(E)
\eeqar
where $X_i=A_i Y_i$, $Y_i=n_i/n_b$, and $\sum X_i=1$.  Let $x_i$ to be the ionization fraction of nuclei $i$, then $Y_{e}=\sum x_i Z_i Y_i= \avg{x Z/A} \approx \avg{x}/2$.  For a fully ionized shell of pure \ni56, we obtain a ratio $n_{^{56}{\rm Ni}}/n_e= X_{^{56}{\rm Ni}}/(A_{^{56}{\rm Ni}}Y_e)\approx 2/A_{^{56}{\rm Ni}}$. $S_0$ is the source function when the line is not broadened.

If $S_{E}$ is constant, from the radiative transfer equation $dI_{E}/d\tau=-I_{E}+S_{E}$, we can get the {intensity $I_{E}$} to be 
\beqar
\label{eq:intensity}
I_{E}(\tau)&=&\frac{dN}{dEdtdAd\Omega}
\nonumber\\
&=&I_{E}(0)e^{-\tau}+S_{E} \int_{0}^{\tau}e^{-(\tau-\tau')}d\tau'
\nonumber\\
&=&I_{E}(0)e^{-\tau}+S_{E}(1-e^{-\tau})
\eeqar
where optical depth $\tau=\int n_{\rm e}(r)\sigma dl=n_{\rm e}\sigma a (l/a)=\tau_a l/a$, {l is the photon path-length}.

Let $\theta$ to be the angle between line of sight and the line between ejecta center and observer, $sin\alpha=D\theta/r$.
Thus total flux from the ejecta should be an integral of the intensity over the solid angle extended by the ejecta, i.e., 
\beqar
\label{eq: flux}
F_{E}&=&\int I_{E} cos\theta d\Omega \approx 2\pi \int I_{E} \theta d\theta
\\
F&=&\int F_{E}dE
\eeqar

For a shell plus core model ({or helium cap model}, shortened as shell model thereafter), motivated by SN2014J \citep{Diehl2014} and ignoring the Compton continuum emission, we assume total $^{56}{\rm Ni}$ mass $M_{^{56}\rm Ni,total}=0.5\msol$ with $10\%$ of the mass distributed at the outmost shell of the ejecta with radius a and depth h, while the remain Ni is in the core {with radius R$_0=(M_{^{56}\rm Ni,core}/M_{\rm ej})^{1/3}a$} and subtended angle $\theta_0\sim R_0/D$, where D is the distance between the earth and the SNIa. The angle subtended by the shell is between $\theta_1\sim (a-h)/D$ and $\theta_2\sim a/D$. Thus the ejecta emission is in the region between $\pm\theta_2$. Sketch of the shell model see Fig.~\ref{fig:shell}.

For $\theta_1\leq\theta\leq\theta_2$, the photon path-length is $l_2=2\sqrt{a^2-(D\theta)^2}$, the intensity is 
\beqar
\label{eq:I2}
I_{E}^2=S_{E}(1-e^{-\tau_a l_2/a}).
\eeqar

For $\theta_0\leq\theta\leq\theta_1$, photon will travel two different regions through the line of sight, $l_1=\sqrt{a^2-(D\theta)^2}-\sqrt{(a-h)^2-(D\theta)^2}$, $s_1=2\sqrt{(a-h)^2-(D\theta)^2}$, the intensity is 
\beqar
\label{eq:I1}
I_{E}^1
&=&S_{E}(1-e^{-\tau_a l_1/a})(e^{-\tau_a (s_1+l_1)/a}+1).
\eeqar

For $0\leq\theta\leq\theta_0$, photon will travel three different regions through the line of sight, $l_0=2\sqrt{R_0^2-(D\theta)^2}, s_0=\sqrt{(a-h)^2-(D\theta)^2}-\sqrt{R_0^2-(D\theta)^2}$, the intensity is 
\beqar
\label{eq:I0}
I_{E}^0
& = & S_{E}(1-e^{-\tau_a l_1/a}) e^{-\tau_a (2s_0+l_0+l_1)/a}
\nonumber\\
& & +S_{E}(1-e^{-\tau_a l_0/a}) e^{-\tau_a (s_0+l_1)/a} 
\nonumber\\
& & +S_{E}(1-e^{-\tau_a l_1/a}).
\eeqar

Therefore, the flux for shell model is 
\beqar
\label{eq:F_shell}
F_{E}^{\rm shell}(E,t)
&=&2\pi (\int_{0}^{\theta_0} I_{E}^0 \theta d\theta+\int_{\theta_0}^{\theta_1} I_{E}^1 \theta d\theta
\nonumber\\
& & +\int_{\theta_1}^{\theta_2} I_{E}^2 \theta d\theta),
\\
F_{\rm shell}(E_i,t)&=&\int_{\rm broadened-E_i} F_{E}^{\rm shell}(E,t) dE.
\eeqar

Similarly, for a core only model where all the Ni is distributed in the core with radius $R_0$ ($h=0$ in the shell model case), the flux is  
\beqar
\label{eq:F_core}
F_{E}^{\rm core}(E,t)
&=&2\pi  (\int_{0}^{\theta_0} I_{E}^0 \theta d\theta+\int_{\theta_0}^{\theta_1} I_{E}^1 \theta d\theta),
\\
F_{\rm core}(E_i,t)&=&\int_{\rm broadened-E_i} F_{E}^{\rm core}(E,t) dE.
\eeqar
with $\theta_1=\theta_2=a/d$.

\section{Lightcurves}
\label{sec: lightcurves}


{As discussed in Section.\ref{sec: methods}, we could identify the SNIa signal and estimate the detection timescales by comparing the SNIa light curves with the background of GBM and BAT detectors .}

\begin{figure}
\centering
\includegraphics[width=0.95\columnwidth]{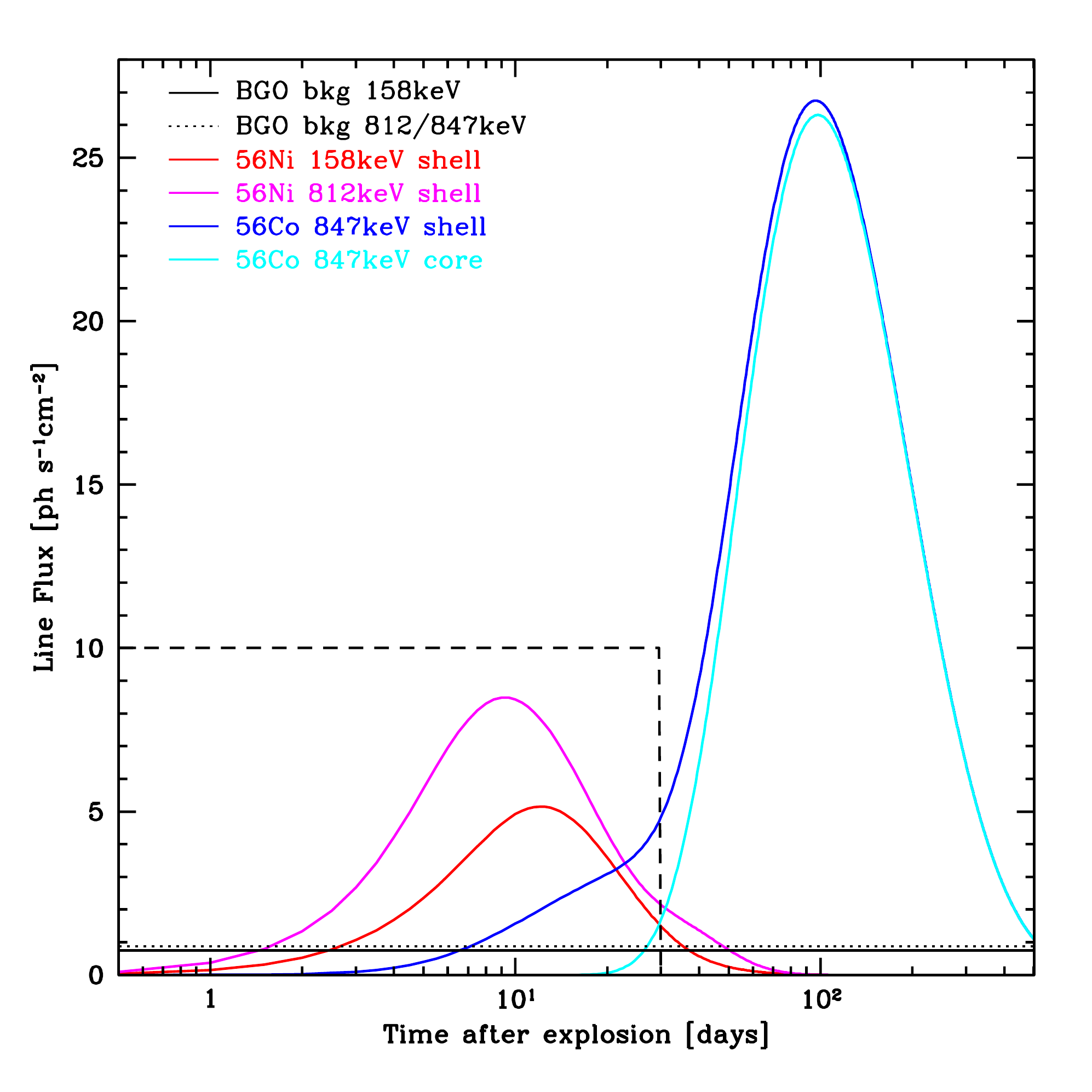}
\nonumber\\
\includegraphics[width=0.95\columnwidth]{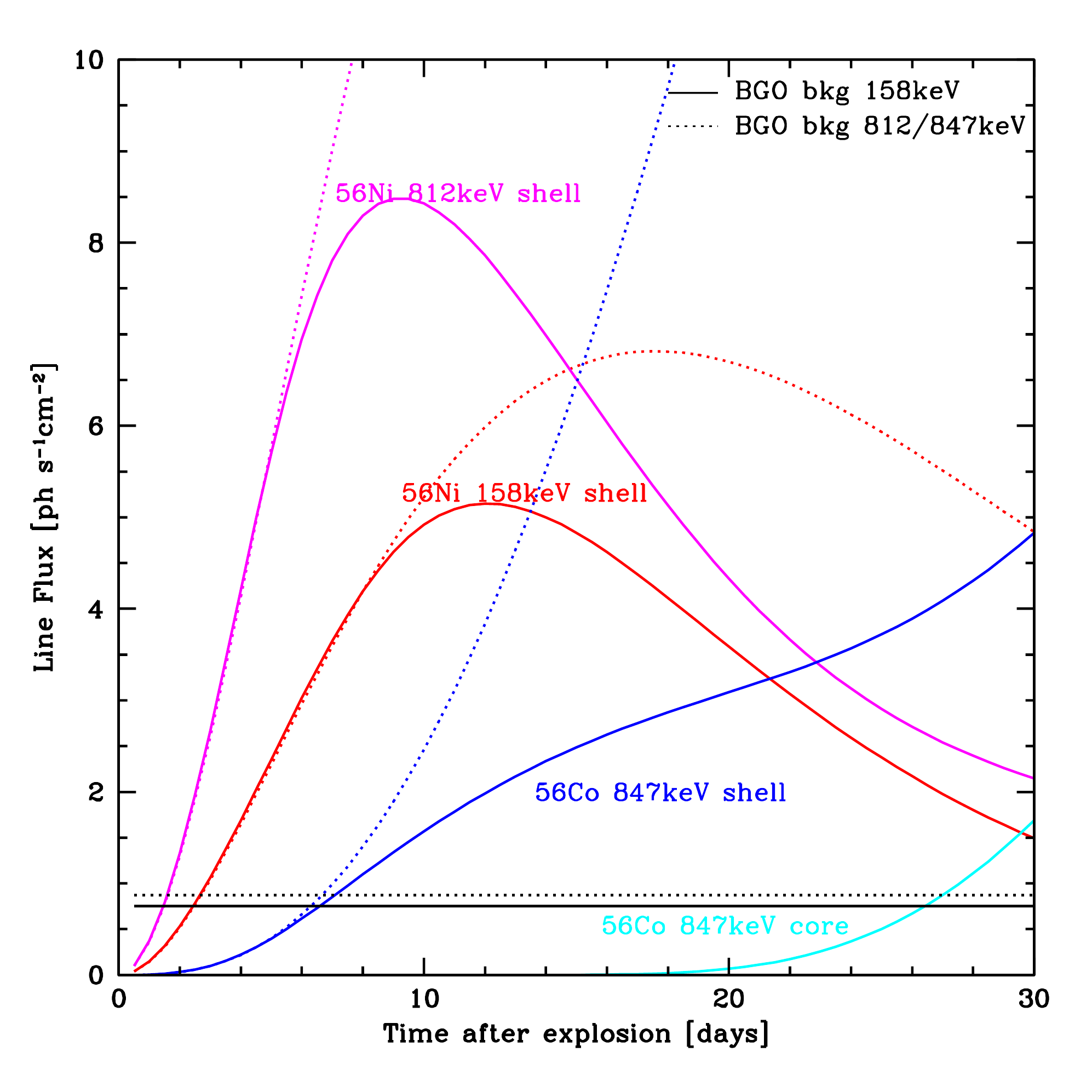}
\caption{
Simulated light curves of the gamma-ray lines from a 10 kpc Galactic SNIa. Black solid line is the average background flux measured in 158 keV bin of the BGO detector, black dotted line is background in 812/847 keV bin. Solid color lines are the simulation results, while dotted color lines are the analytical optically thick results (Eq.\ref{eq:f1}). Red lines are the $^{56}{\rm Ni}$ 158 keV line signal intensity variation for the shell model, and magenta for the {812} keV lines. Blue lines are the $^{56}{\rm Co}$ 847 keV line signal intensity variation for the shell model and cyan is the 847 keV line for the core model. Upper panel: Light curves ranging in 500 days. Lower panel: Light curves ranging in 30 days, which is the black dashed region in left panel.
\label{fig:light_curve}
}
\end{figure}

Fig.~\ref{fig:light_curve} shows the simulated SNIa light curves of $^{56}{\rm Ni}$ decay lines (the dominant 158 keV and 812 keV lines) and $^{56}{\rm Co}$ decay line (the dominant 847 keV line), for both the shell model and core model.  
For the optimistic case (shell model), we can see that the line fluxes from $^{56}{\rm Ni}$ decay will exceed the BGO background at first few days after the explosion and reach the peak at $\sim$10 days when the fluxes are $\gtrsim$4-5 times higher than the background. 
Therefore, if the shell model is true, the $^{56}{\rm Ni}$ decay signal from a Galactic SNIa will be noticed as early as first days after the explosion. 
Although $^{56}{\rm Co}$ decay line fluxes are much smaller compared to $^{56}{\rm Ni}$ at early days, they still exceed the BGO background flux at $\sim$ 8 days and reach the peak at $\sim$100 days. 

For the most pessimistic case (core model), the ejecta remains optically thick at first tens of days after the explosion, thus $^{56}{\rm Ni}$ lines will be too weak to be observed by the detector. Also $^{56}{\rm Co}$ lines will emerge much later than the shell model case, after $\sim$20 days. Then $^{56}{\rm Co}$ lines will reach the peak at a similar time and evolve to be the same as the shell model at later time, with the same total $^{56}{\rm Ni}$ mass.

Similar conclusions will be obtained for the $^{56}{\rm Ni}$ 158keV line observed by BAT detectors with a similar background flux at the 158keV bin.
In addition, the upper panel in Fig.~\ref{fig:light_curve} show that the analytical formula Eq. (1) fits the simulated light curves perfectly at the first few days when the ejecta is optically thick, meaning that we could predict accurately the line signal from a Galactic SNIa at early days after the explosion and the observations will give important information about the ejecta.

\begin{figure}
\centering
\includegraphics[width=0.95\columnwidth]{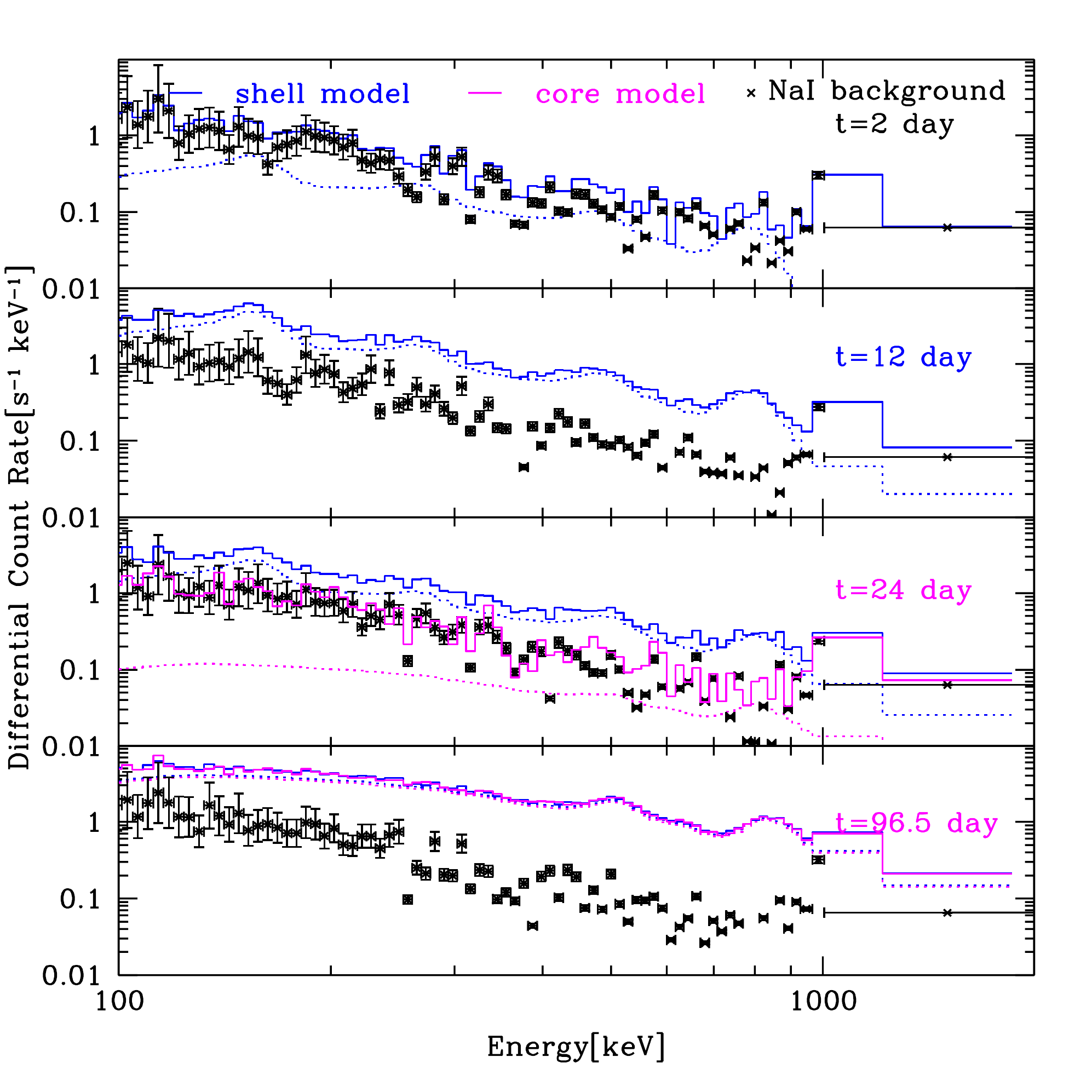}
\caption{
Simulation of $^{56}{\rm Ni}$ and $^{56}{\rm Co}$ gamma-ray lines 
signal seen in a GBM NaI detector, from a 10 kpc Galactic SNIa at day2 (above), day12 (middle above), day24 (middle below) and day96.5 (below) after explosion. The black points represent a typical background spectrum from detector n3. Dotted lines are the simulated SNIa \ni56 and \co 56 decay lines signal, and solid lines are the total signal expected to be seen by the detectors (sum of the background and SNIa signal). Blue lines are for shell model, magenta are for core model.   
\label{fig:NaI_spectra}
}
\end{figure}

\section{Localization}
\label{sec: localization}

Because {\em Fermi} orbits the Earth with an altitude of 555 km $\sim$$10\% R_\oplus$, about $30\%$ of {\em Fermi}'s field of view is always blocked by the Earth.  About $85\%$ of the sky is occulted in one orbit, so point sources will typically be eclipsed once per orbit, as seen in Fig.~\ref{fig:location}. From the eclipse flux decrement and timing, sources can be identified and located.  The Earth Occultation Technique and Earth Occultation Imaging have been successfully used to search for known and unknown point sources for GBM \citep{Wilson-Hodge2012, Rodi}. The occultation duration will depend on the elevation angle $\beta$ between the source and the orbital plane. The angular resolution ranges from $\sim$$0.5^{\circ}$ for $\beta=0$ to $\sim$$1.25^{\circ}$ for $\beta=66^{\circ}$, and the timescale of localization is $\sim$P, where $P=96$ minutes is {\em Fermi}'s orbital period. For $\beta > 66^{\circ}$, occultation does not occur, GBM need to wait to precess to another orbit to have signal blocked. 

BAT may be the first, perhaps the {\em only} way to detect a Galactic SNIa simultaneously in 
both X-ray and UV and optical bands. 
{\it Swift}/BAT has PSF of 22.5 arcmin for an on-axis source \citep{BAT1}, 
this alone is good enough to be easily within the field of view for many optical/IR surveys, 
such as the {\em Dark Energy Survey} and later {\em LSST}.

\begin{figure}
\centering
\includegraphics[width=0.95\columnwidth]{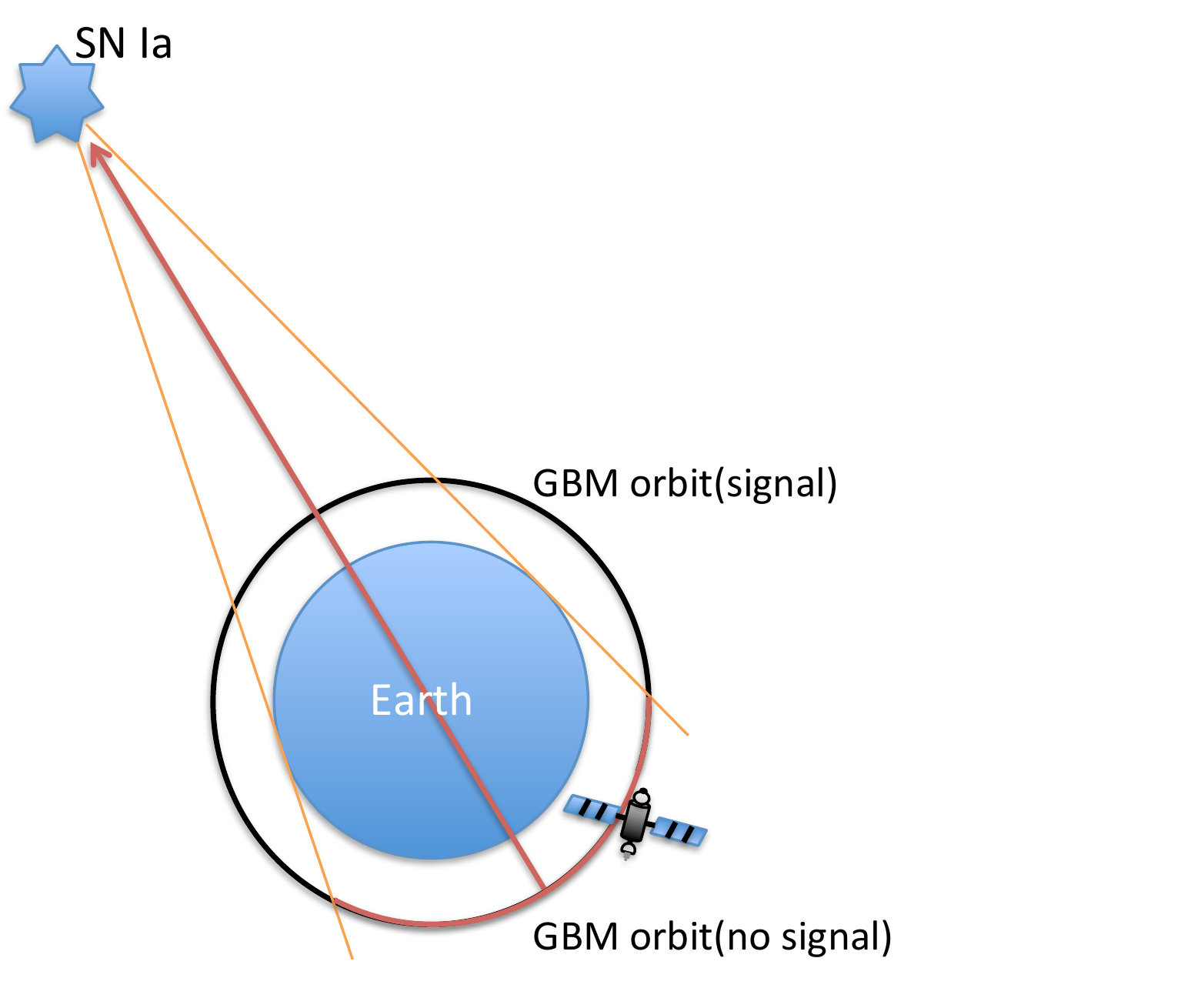}
\caption{
Sketch of how the Earth occultation technique localizes a Galactic SNIa. The signal will be occulted by the Earth during one orbit of the satellite, and the source position can be found by associating the occulting time and the satellite position.
\label{fig:location}
}
\end{figure}

A Galactic SNIa--once identified as such--would likely be observable across the electromagnetic spectrum, including at wavelengths not yet seen in outburst. 
If there is an X-ray signal, {\em Swift}/XRT precision is
good to about 1 pixel or $\sim$2.4 arcsec.
If {\em Swift}/UVOT sees a signal, the precision is similar, which will be in the FOV of all optical/IR telescopes.
But UVOT is difficult to localize the Galactic SNIa due to the crowdedness of stars and bright sources in the Galactic plane. Therefore after BAT discover the Galactic SNIa, XRT will followup to confirm the SNIa and localize it with a relatively cleaner background. If there is no X-rays seen, UVOT will be used to localize the SNIa with white filter, finally take the spectra with both UV and optical filters. 
{After {\em Swift} localization, a Galactic SNIa merits repeated followup at all wavelengths.  At high energies, {\em NuSTAR}, {\em INTEGRAL}, and {{\em RHESSI}-like telescopes} can probe observations of both nuclear lines and continuum.
No SNIa has been seen in radio at early times, so such observations would offer powerful new constraints or new detection \citep[e.g.,][]{radio}, and possibly allow for 
early interferometric imaging.
Obviously, optical/IR followup will be of great value as well{\footnote{Considering about the Galactic absorption, using IR telescopes like {\em Spitzer} and later {\em JWST} for next followup observation will be valuable. The SNIa infrared light curves generally have two maxima separated by $\sim$20-30 days \citep[e.g.,][]{Johansson}, the second maximum observation will benefit the diagnostic of the physical properties of the ejecta, like the degree of \ni56 mixing \citep{IR}.}}.
When BAT will find the SNIa depends on when it happens to be in its FOV, and thus on the {\em Swift} scan strategy.
}



\bsp	
\label{lastpage}
\end{document}